\documentclass[12pt]{elsart}
\usepackage{amssymb}

\usepackage{epsfig}
\usepackage{fleqn}

\begin{document}

\hoffset=0.5cm 
\begin{frontmatter}
\begin{flushright}
HIP-1999-78/TH \\
PPD-IOP-99/25
\end{flushright}

\vspace{1 cm}

\title{Large hierarchy from extra dimensions}

\author{Masud Chaichian\thanksref{em1}}
\address{Department of Physics, High Energy 
Physics Division, University of Helsinki and 
Helsinki Institute of Physics, P.O. Box 9 (Siltavuorenpenger 20 C), \\
FIN-00014 Helsinki, Finland }
\author{Archil B. Kobakhidze\thanksref{em2}}
\address{Andronikashvili Institute of Physics, Tamarashvili str. 6, \\
GE-380077 Tbilisi, Georgia}
\thanks[em1]{E-mail: Masud.Chaichian@helsinki.fi. Supported by the Academy 
of Finland under the Project No. 163394.}
\thanks[em2]{E-mail: achiko@iph.hepi.edu.ge. Partially supported by the Grant No.
2-10 of the Georgian Academy of Sciences, by the Georgian Young Scientists 
Presidential Award and by the INTAS Grant 96-155.}

\begin{abstract}
We argue that the familiar gauge hierarchy between the fundamental Planck 
scale $M_{Pl}$ and the electroweak scale 
$M_{W}$, can be naturally explained 
in higher dimensional theories with relatively large radii 
($R_c > 1/M_{Pl}$) extra dimensions.  
In particular, we show that it is possible that the electroweak Higgs mass at high 
energies is of the order of  $M_{Pl}$, but radiative corrections drive it 
to an infrared stable fixed--point $\sim M_{W}$ at low energies thus inducing a 
large hierarchy without any fine tuning of parameters.
\end{abstract}
\end{frontmatter}

\newpage

Despite the fact that the Standard Model (SM) of strong and electroweak
interactions successfully describes all observed phenomena at currently
accessible energies $M_{W}\sim 100$ GeV, there are several indications for
new physics beyond the SM to exist. From the theoretical point of view,
perhaps the most strict hint for such new physics is the gauge hierarchy
problem which deals with the natural explanation not only of the hierarchy
of scales, $M_{W}/\Lambda \ll 1$, but also of the stability of this
hierarchy under the quantum corrections, where the SM is regarded as a
low--energy limit of a certain more fundamental theory with characteristic
scale $\Lambda $ ($\Lambda $ can be, say, the grand unified scale $%
M_{GUT}\sim 10^{16}$ GeV, or the Planck scale $M_{Pl}\approx 10^{18}$ GeV).
The solution to this problem in four dimensions has been usually attributed
to supersymmetric models, or models with dynamical gauge symmetry breaking
with characteristic scales of the order of TeV.

On the other hand, it is widely believed that a unified theory of all
fundamental forces in nature (including gravity) can be consistently
constructed in space-time with more than four dimensions. That is why the
novel attack to the gauge hierarchy problem within the higher dimensional
theories, much motivated by studies of non--perturbative superstring and
M--theories, has recently received considerable attention. It has been
proposed in \cite{AHDD1} that the fundamental Planck scale can be as low as
few TeV, requiring the presence of two extra compact dimensions at least
with sub--millimeter radii, $R_{c}=\frac{1}{\mu _{c}}\sim 1$ mm, and that SM
particles and forces are confined on a 3--brane in the extra dimensions in
order to avoid the conflict with presently accessible experiments. However,
while this scenario does eliminate the hierarchy between the weak $M_{W}$
and Planck scale $M_{Pl}$, it introduces a new hierarchy between $\mu _{c}$
and $M_{W}$ ($\frac{\mu _{c}}{M_{W}}\approx \frac{M_{W}}{M_{Pl}}$) and
stability of large extra dimensions also remains as a critical question. In
this connection, another distinct higher dimensional scenario proposed in 
\cite{RS1} provides an exponential hierarchy between the weak $M_{W}$ and
Planck $M_{Pl}$ scales through the special solutions for the
five--dimensional Einstein equation with the cosmological constant and
3--branes involving ''warp'' factor in the metric tensor \cite{warp}%
\footnote{%
A remarkable thing offered by such a solutions is the possibility to
reproduce the four--dimensional Newton's law in our universe even with
infinitely large (non--compact) extra dimensions. However, to solve the
gauge hierarchy problem in this case more complex constructions are required 
\cite{NCGH}.}. Then the required hierarchy is naturally generated even with
small extra dimensions ($\frac{\mu _{c}}{M_{Pl}}\approx 10^{-2}$), so the
model easily avoids current constraints from particle physics \cite{cons1},
astrophysics and cosmology \cite{cons2}\footnote{%
It has been emphasized in \cite{cosm} the troubles with cosmological
expansion in the scenario where the SM is placed on the 3--brane with
negative tension \cite{RS1}. Also, the difficulty to reproduce gauge
coupling unification is pointed out in \cite{UNIF}.}. Several studies of
rich phenomenological and cosmological consequences of such a scenario have
been proposed recently \cite{WARPPHEN}. Note, however, that a common
deficiency of the presented models is the \textit{ad hoc }fine tunings
required between the cosmological constants in the bulk and on the
3--branes, in order to obtain the desired solution to the Einstein
equations. The stability of such fine tuning seems also to be problematic.

In this paper we suggest a new mechanism for gauge hierarchy in higher
dimensional theories with relatively large radii compact extra dimensions.
We explore the old idea that the large hierarchy of scales can be explained
without any fine tuning of parameters if the electroweak Higgs boson
anomalous mass dimension is $A\geq 2$ \cite{BW}\footnote{%
Somewhat similar idea have been proposed in \cite{GUTPAT} to study the
patterns of GUT symmetry breaking.}. However, $A\geq 2$ is highly
undesirable, since anomalous mass dimension $A$, being proportional to
coupling constants, is usually $\ll 1$, unless some of the couplings (Higgs
self--interacting coupling or/and Yukawa couplings) are non--perturbative
below the scale $\Lambda $, or there is an unrealistically large number of
degrees of freedom ensuring $A\geq 2$ \cite{BW}. Needless to say, that it is
very difficult (if ever possible) to construct a realistic model obeying
such conditions. In higher dimensional theories, however, the situation is
drastically changed. The point is that, owing to the power--law (in contrast
to the logarithmic in four dimensions) evolution of the theory parameters,
the Higgs vacuum expectation value (VEV), while being of the order of $%
\Lambda $ at high energies, rapidly decrease down to the infrared stable
fixed point $\sim M_{W}$ even for the small values of $A$, thus naturally
inducing a large hierarchy even in the case of SM with the ordinary number
of colours and flavours. Let us demonstrate this phenomenon explicitly on
the simplified example of the $SU(N)$--symmetric Higgs--Yukawa system with $%
N_{c}$ colours.

Our starting action in $D=4+\delta $ dimensions ($\delta $ is the number of
extra compact dimensions) is 
\begin{eqnarray}
S_{\Lambda _{0}} &=&\int d^{4+\delta }[Z(\Lambda _{0})\partial _{\mu }\phi
^{+}\partial ^{\mu }\phi -\mu ^{2}(\Lambda _{0})\phi ^{+}\phi +\frac{1}{2}%
\lambda (\Lambda _{0})\left( \phi ^{+}\phi \right) ^{2}  \nonumber \\
&&+Z_{L}(\Lambda _{0})\overline{\psi }_{L}i\gamma _{\mu }\partial ^{\mu
}\psi _{L}+Z_{R}(\Lambda _{0})\overline{\psi }_{R}i\gamma _{\mu }\partial
^{\mu }\psi _{R}  \nonumber \\
&&+\left( h(\Lambda _{0})\overline{\psi }_{L}\phi \psi _{R}+h.c.\right) ],
\label{1}
\end{eqnarray}
where $\phi $ is an $N$--component complex scalar field, $\psi _{L}$ is an $%
N $--component left--handed fermion field with $N_{c}$ colours and $\psi
_{R} $ is a right--handed $SU(N/2)$--singlet fermion with $N_{c}$ colours
again. $Z$, $Z_{L}$, and $Z_{R}$ in (\ref{1}) are the field renormalization
factors which we choose to be equal to 1 at the scale $\Lambda _{0}$. In the
case of $N=2$ and $N_{c}=3$ the action (\ref{1}) is just the SM action in
the limit of vanishing gauge couplings and Higgs--Yukawa couplings except
for one type of quarks.

Theory with action (\ref{1}) in higher dimensions ($\delta \neq 0$) is known
to be non--renormalizable, but it can be well defined by introducing an
ultraviolet cut--off $\Lambda _{0}$, which is natural to identify with the
fundamental Planck scale $M_{Pl}$. At low energies one can consistently
describe the theory using Wilsonian effective action approach \cite{Wilson}%
\footnote{%
Recently, this approach have been applied to the higher dimensional theories
in \cite{KUBO}.}. The basic idea behind this approach is first to integrate
out momentum modes between a cut--off scale $\Lambda _{0}$ and lower energy
scale $\Lambda $, rather than to integrate over all momentum modes in one
go. The remaining integral from $\Lambda $ to zero may again be expressed as
a partition function, but the bare action $S_{\Lambda _{0}}$ (\ref{1}) is
replaced by a complicated effective action $S_{\Lambda }$ (Wilsonian
effective action) and the overall cut--off $\Lambda _{0}$ by the effective
cut--off $\Lambda $, in such a way that all physics, i.e. Green functions,
are left invariant. The difference in $S_{\Lambda }$ induced by the change
of the cut--off is determined integrating ''shell modes'' with momenta
between $\Lambda $ and $\Lambda +\delta \Lambda $ and for an infinitesimal $%
\delta \Lambda $ becomes a Gaussian path integral which can be exactly
carried out. Thus, the scale dependence of the Wilsonian effective action is
given by the exact functional differential equation 
\begin{equation}
\Lambda \frac{\partial S_{\Lambda }}{\partial \Lambda }=\mathcal{O}%
[S_{\Lambda }]\ ,  \label{2}
\end{equation}
where $\mathcal{O}[S_{\Lambda }]$ is a non--linear operator acting on the
functional $S_{\Lambda }$. However, for the practical calculations it is
inevitable to approximate the evolution equation (\ref{2}). Usually, one
expands the effective action $S_{\Lambda }$ in terms of the number of
derivatives included in the general operators. In the first order of this
approximation any derivative couplings are dropped except for the kinetic
term. This is the so--called local potential approximation (LPA) \cite{LPA},
which we adopt here. In the ''sharp cut--off'' scheme we obtain the
renormalization group (RG) equation for the local effective scalar potential
(Wegner--Houghton equation): 
\begin{eqnarray}
\Lambda \frac{\partial V}{\partial \Lambda } &=&-\frac{K_{D}}{2}\Lambda
^{D}[\ln \left( Z\Lambda ^{2}+V^{^{\prime }}(\rho )+2\rho V^{^{\prime \prime
}}(\rho )\right) +  \nonumber \\
&&(2N-1)\ln \left( Z\Lambda ^{2}+V^{^{\prime }}(\rho )\right) -2^{\frac{D}{2}%
}N_{c}\ln \left( Z_{L}Z_{R}\Lambda ^{2}+h^{2}\rho \right) ],  \label{3}
\end{eqnarray}
where $\rho =\frac{1}{2}|\phi |^{2}$, prime denotes the derivative with
respect to $\rho $ and $K_{D}$ is the $D$--dimensional angular integral 
\begin{equation}
K_{D}=\frac{2^{1-D}}{\pi ^{-D/2}\Gamma (D/2)}  \label{4}
\end{equation}
($\Gamma $ is the Euler function). Now, since we are interested in the
spontaneously broken regime, let us expand the effective potential $V$
around its minimum $\rho _{0}(\Lambda )$ (commoving scheme) \cite{CMS}: 
\begin{equation}
V(\rho )=u_{0}(\Lambda )+u_{1}(\Lambda )\left( \rho -\rho _{0}(\Lambda
)\right) +\frac{1}{2}u_{2}(\Lambda )\left( \rho -\rho _{0}(\Lambda )\right)
^{2}+...,  \label{5}
\end{equation}
where the minimum $\rho _{0}(\Lambda )$ is defined from the extremum
equation 
\begin{equation}
\left. \frac{\partial V}{\partial \rho }\right| _{\rho =\rho _{0}}=0\ .
\label{6}
\end{equation}
Taking the total $\ln \Lambda $--derivative of (\ref{6}), we obtain RG
equation for $\rho _{0}(\Lambda )$:

\begin{eqnarray}
\Lambda \frac{d\rho _{0}}{d\Lambda } &=&\frac{K_{D}}{2}\Lambda ^{D}[\frac{3}{%
Z\Lambda ^{2}+2\rho _{0}\lambda }+(2N-1)\frac{1}{Z\Lambda ^{2}}-  \nonumber
\\
&&2^{\frac{D}{2}}N_{c}\frac{h^{2}}{\lambda }\frac{1}{Z_{L}Z_{R}\Lambda
^{2}+h^{2}\rho _{0}}],  \label{7}
\end{eqnarray}
where we have denoted the scalar quadric coupling as $\lambda (\Lambda
)\equiv u_{2}(\Lambda )=\left. \frac{\partial ^{2}V}{\partial \rho ^{2}}%
\right| _{\rho =\rho _{0}}$. By substituting (\ref{5}) into (\ref{3}), it
will be also easy to obtain an infinite set of coupled RG equations for $%
u_{i}(\Lambda )=$ $\left. \frac{\partial ^{i}V}{\partial \rho ^{i}}\right|
_{\rho =\rho _{0}}$($i=3,4,...$) couplings, 
\begin{equation}
\Lambda \frac{du_{i}}{d\Lambda }=u_{i+1}\Lambda \frac{d\rho _{0}}{d\Lambda }%
+\left. \frac{\partial ^{i}\left( \Lambda \frac{\partial V}{\partial \Lambda 
}\right) }{\partial \rho ^{i}}\right| _{\rho =\rho _{0}}\ .  \label{8}
\end{equation}
Thus, the Wilson RG framework allows, in principle, to take into account
higher--dimensional operators in the evolution of couplings in (\ref{5}).
However, we restrict ourself to consider only the potential in the vicinity
of $\rho _{0}(\Lambda )$ approximated by a $\phi ^{4}$--potential. This
corresponds to the truncation 
\begin{equation}
u_{3}=u_{4}=...=0\ .  \label{9}
\end{equation}
In this approximation the evolution equation for the scalar quadric coupling 
$\lambda (\Lambda )$ looks as 
\begin{eqnarray}
\Lambda \frac{d\lambda }{d\Lambda } &=&\frac{K_{D}}{2}\Lambda ^{D}[\frac{%
9\lambda ^{2}}{(Z\Lambda ^{2}+2\rho _{0}\lambda )^{2}}+(2N-1)\frac{\lambda
^{2}}{Z^{2}\Lambda ^{4}}-  \nonumber \\
&&2^{\frac{D}{2}}N_{c}\frac{h^{4}}{(Z_{L}Z_{R}\Lambda ^{2}+h^{2}\rho
_{0})^{2}}]\ .  \label{10}
\end{eqnarray}
Now, let us define the renormalized quantities 
\begin{eqnarray}
\rho _{R} &=&Z\rho _{0}\ ,  \nonumber \\
\lambda _{R} &=&Z^{-2}\lambda \ ,  \nonumber \\
h_{R}^{2} &=&Z_{L}^{-1}Z_{R}^{-1}Z^{-1}h^{2}\ ,  \label{11}
\end{eqnarray}
and the anomalous dimensions 
\begin{eqnarray}
\gamma &=&-\Lambda \frac{d\ln Z}{d\Lambda }\ ,  \nonumber \\
\gamma _{L} &=&-\Lambda \frac{d\ln Z_{L}}{d\Lambda }\ ,  \nonumber \\
\gamma _{R} &=&-\Lambda \frac{d\ln Z_{R}}{d\Lambda }\ .  \label{12}
\end{eqnarray}
The anomalous dimensions (\ref{12}) had been calculated in \cite{ANDIM} in
the smooth cut--off approximation (average action), which results we adopt
here: 
\begin{eqnarray}
\gamma &=&\frac{K_{D}}{D}\Lambda ^{D-4}2^{\frac{D}{2}}N_{c}h_{R}^{2}\ , 
\nonumber \\
\gamma _{L} &=&\frac{2K_{D}}{D}\Lambda ^{D-4}2^{\frac{D}{2}}h_{R}^{2}\ , 
\nonumber \\
\gamma _{R} &=&\frac{2K_{D}}{D}\Lambda ^{D-4}2^{\frac{D}{2}}Nh_{R}^{2}\ .
\label{13}
\end{eqnarray}
Then, from Eqs. (\ref{7}) and (\ref{10})--(\ref{13}) the RG equations for
renormalized parameters $\rho _{R}$ and $\lambda _{R}$ can be easily
obtained, 
\begin{equation}
\Lambda \frac{d\rho _{R}}{d\Lambda }=\frac{K_{D}}{2}\left( (2N+2)-2^{\frac{D%
}{2}}N_{c}\frac{h_{R}^{2}}{\lambda _{R}}\right) \Lambda ^{D-2}+A_{\delta
}\rho _{R}\ ,  \label{14}
\end{equation}
with 
\begin{equation}
A_{\delta }=-\frac{K_{D}}{2}\Lambda ^{D-4}\left[ 6\lambda _{R}-2^{\frac{D}{2}%
}N_{c}\frac{h_{R}^{4}}{\lambda _{R}}\right] -\gamma  \label{15}
\end{equation}
and 
\begin{equation}
\Lambda \frac{d\lambda _{R}}{d\Lambda }=\frac{K_{D}}{2}\Lambda
^{D-4}[(2N+8)\lambda _{R}^{2}-2^{\frac{D}{2}}N_{c}h_{R}^{4}]+2\gamma \lambda
_{R}\ ,  \label{16}
\end{equation}
where we have expanded the right--hand sides of equations (\ref{14}) and (%
\ref{16}) in powers of $\frac{2\rho _{R}\lambda _{R}}{\Lambda ^{2}}$ and $%
\frac{h_{R}^{2}\rho _{R}}{\Lambda ^{2}}$ and have kept only the leading
terms $O(\rho _{R})$ in (\ref{14}) and $O(\lambda _{R}^{2})$ and $%
O(h_{R}^{4})$ in (\ref{16}). This approximation is valid if 
\begin{eqnarray}
\frac{2\rho _{R}\lambda _{R}}{\Lambda ^{2}} &\ll &1\ ,  \nonumber \\
\frac{h_{R}^{2}\rho _{R}}{\Lambda ^{2}} &\ll &1\ ,  \label{17}
\end{eqnarray}
and, as we will show below, correctly reproduce the results of the standard
one--loop perturbation theory. Actually, the relations $2\rho _{R}\lambda
_{R}(m_{S})=m_{S}^{2}$ and $h_{R}^{2}\rho _{R}(m_{F})=m_{F}^{2}$ determine
the physical masses of the scalar radial mode and the fermion field,
respectively, and thus below the scales $m_{S}$ and $m_{F}$ scalar and
fermion fields become decoupled correspondingly.

$A_{\delta }$ of Eq. (\ref{15}), which appears in the last term of (\ref{14}%
) is just the anomalous dimension for the VEV of the scalar field. Clearly, $%
A_{\delta }$ has to be evaluated at the critical value $\rho _{R}^{cr}$,
which is defined by solving (\ref{14}) with first term only 
\begin{equation}
\rho _{R}^{cr}(\Lambda \rightarrow 0)=0\ .  \label{18}
\end{equation}
The critical value $\rho _{R}^{cr}$ corresponds to the phase transition
between the symmetric and spontaneously broken phases. Thus the anomalous $%
A_{\delta }$ has a simple physical interpretation \cite{INTER}: It governs
the scale dependence of the deviation of scalar field VEV from the critical
value (\ref{18}) resulting to the nonzero VEV for the scalar field 
\begin{equation}
\left\langle \phi ^{2}\right\rangle =\lim_{\Lambda \rightarrow 0}\rho
_{R}(\Lambda )=\lim_{\Lambda \rightarrow 0}(\rho _{R}^{cr}(\Lambda
)+v_{R}^{2}(\Lambda ))=v^{2}\ ,  \label{19}
\end{equation}
\begin{equation}
\Lambda \frac{dv_{R}^{2}}{d\Lambda }=A_{\delta }v_{R}^{2}\ .  \label{20}
\end{equation}

The evolution equation for the renormalized Yukawa coupling $h_{R}$ can be
also derived from the effective action $S_{\Lambda }$ with nonzero fermion
background fields $\psi $. Keeping the leading $O(h_{R}^{4})$ terms one
obtains 
\begin{equation}
\Lambda \frac{dh_{R}^{2}}{d\Lambda }=(\gamma +\gamma _{L}+\gamma
_{R})h_{R}^{2}\ .  \label{21}
\end{equation}

It is obvious that the set of the RG equations (\ref{16}), (\ref{20}) and (%
\ref{21}) is approximately valid for $\Lambda \gg \mu _{c}$ if some spatial
dimensions are compactified. Here we simply ignore the threshold effects and
assume that below the compactification scale $\mu _{c}$ the theory is
four--dimensional, while above $\mu _{c}$ it is $D$--dimensional with flat
(non--compact) $\delta $ extra dimensions. In $D$ dimensions one can define
effective four--dimensional parameters by appropriately rescaling the fields
and couplings\footnote{%
Here we assume that $\delta =D-4$ extra dimensions are compactified on a
circle of a fixed radius $R_{c}=\frac{1}{\mu _{c}}$. The factor $(2\pi
R_{c})^{\delta }$ is just the volume of extra space appeared in the
effective four--dimensional action after one integrates over the extra space.%
}: 
\begin{eqnarray}
\overline{v}^{2} &=&(2\pi R_{c})^{\delta }v_{R}^{2}\ ,  \nonumber \\
\overline{\lambda } &=&(2\pi R_{c})^{-\delta }\lambda _{R}\ ,  \nonumber \\
\overline{h} &=&(2\pi R_{c})^{-\frac{\delta }{2}}h_{R}\ .  \label{22}
\end{eqnarray}
For the four--dimensional renormalized parameters (\ref{22}) we finally
obtain the evolution equations: 
\newcounter{adeq} \addtocounter{adeq}{1} \renewcommand{\theequation}{%
\arabic{equation}\alph{adeq}} 
\begin{equation}
\Lambda \frac{d\overline{v}^{2}}{d\Lambda }=\frac{K_{D}}{2}\left( \frac{2\pi
\Lambda }{\mu _{c}}\right) ^{\delta }\left[ -6\overline{\lambda }-\frac{2^{%
\frac{D}{2}+1}}{D}N_{c}\overline{h}^{2}+2^{\frac{D}{2}}N_{c}\frac{\overline{h%
}^{4}}{\overline{\lambda }}\right] \overline{v}^{2}\ ,  \label{23a}
\end{equation}
\addtocounter{equation}{-1} \addtocounter{adeq}{1} 
\begin{equation}
\Lambda \frac{d\overline{\lambda }}{d\Lambda }=\frac{K_{D}}{2}\left( \frac{%
2\pi \Lambda }{\mu _{c}}\right) ^{\delta }\left[ (2N+8)\overline{\lambda }%
^{2}+\frac{2^{\frac{D}{2}+2}}{D}N_{c}\overline{h}^{2}\overline{\lambda }-2^{%
\frac{D}{2}}N_{c}\overline{h}^{4}\right] \ ,  \label{23b}
\end{equation}
\addtocounter{equation}{-1}\addtocounter{adeq}{1} 
\begin{equation}
\Lambda \frac{d\overline{h}^{2}}{d\Lambda }=\frac{K_{D}}{2}\left( \frac{2\pi
\Lambda }{\mu _{c}}\right) ^{\delta }\frac{2(N+1)+2^{\frac{D}{2}}N_{c}}{D}%
\overline{h}^{4}\ .  \label{23c}
\end{equation}
\renewcommand{\theequation}{\arabic{equation}} Note, that by taking $\delta
=0$ the set of Eqs. (23a-c) correctly reproduces the familiar one--loop
results of perturbation theory in four dimensions. The crucial role of the
extra dimensions in solving the gauge hierarchy problem can be seen from
Eqs. (23a-c) even without performing numerical calculations. Indeed,
ignoring for the moment the running of $\overline{\lambda }$ and $\overline{h%
}$, one finds from (\ref{23a}) 
\begin{equation}
\frac{\overline{v}(M_{W})}{\overline{v}(M_{Pl})}=\left( \frac{M_{W}}{\mu _{c}%
}\right) ^{\frac{\omega _{0}}{2}}\exp \left[ \frac{(2\pi )^{\delta }}{%
2\delta }\omega _{\delta }\left( 1-\left( \frac{M_{pl}}{\mu _{c}}\right)
^{\delta }\right) \right] \ ,  \label{24}
\end{equation}
where $\omega _{\delta }=A_{\delta }/(\Lambda )^{\delta }$. The exponential
factor in (\ref{24}) can be naturally small in the case of extra dimensions (%
$\delta \neq 0$) even for small (but positive) values of $\omega _{\delta }$%
, providing the desired hierarchy $\frac{\overline{v}(M_{W})}{\overline{v}%
(M_{Pl})}\approx \frac{M_{W}}{M_{Pl}}$, while in four dimensions this ratio
is of the order of $O(1\div 10)$ unless $\omega _{0}=A_{0}\gtrsim 2$, that
can, however, never be obtained in perturbation theory since for small
couplings $A_{0}$ is proportional to these couplings \cite{BW}.

Of course, the actual solutions of the set of Eqs. (23a-c) is more
complicated, since the Yukawa and self--interaction couplings also exhibit
fast (power--law) running and the approximation of the constant $\overline{%
\lambda }$ and $\overline{h}$ is very crude. We have analyzed Eqs. (23)
numerically. The Yukawa coupling $\overline{h}$ rapidly decreases going down
in the energy region between $M_{Pl}$ and $\mu _{c}$ and drives to the
infrared stable fixed--point $\overline{h}=0$. If the Yukawa coupling
dominates over the $\overline{\lambda }$ ($\overline{h}^{2}\gg \overline{%
\lambda }$)\footnote{%
Note that the strong and nearly equal Yukawa couplings at higher energies
provide new explanation for the flavour symmetry breaking in theories with
extra dimensions since they are exponentially suppressed at low energies due
to the renormalization effects \cite{FERM}. Quasi fixed--point solutions of
high dimensional RG equations in connection with fermion mass hierarchies
are thoroughly discussed in \cite{FERM1}} then $\overline{\lambda }$ at the
same time increases for smaller energies, until $\overline{\lambda }$
becomes large enough so that the terms proportional to $\overline{\lambda }%
^{2}$ and $\overline{h}^{2}\overline{\lambda }$ cancel the term proportional
to $\overline{h}^{4}$ in (\ref{23b}). Thus, $\overline{\lambda }$ approaches
the infrared stable fixed--point, $\overline{\lambda }\sim \overline{h}^{2}$%
. At the same time, even starting with large initial values of $\overline{v}%
(M_{Pl})\lesssim M_{Pl}$, $\overline{v}$ rapidly decreases and below the $%
\mu _{c}$ changes very slowly. Thus, for certain $\mu _{c}$ and $\delta $
the mean value of anomalous dimension $A_{\delta }$ can be equal to 2, which
means that $\frac{v^{2}(\Lambda )}{\Lambda ^{2}}$ has an infrared stable
quasi fixed--point. If the parameters are in the vicinity of this
fixed--point, one can naturally explain the hierarchy $\frac{\overline{v}%
(M_{W})}{\overline{v}(M_{Pl})}\approx \frac{M_{W}}{M_{Pl}}$.

Such an approximate fixed--point behaviour of the parameters in our toy
model is expressed in Figs. 1 and 2. We fixed the values of
self--interaction coupling and VEV of the scalar field at 174 Gev to be $%
\overline{\lambda }(174$ GEV$)=0.25$ and $\overline{v}(174$ GEV$)=174$ GeV
and find that small variation of the initial Yukawa coupling $\overline{h}%
(174$ GEV$)=0.55$ (from 0.545 to 0.553; this range corresponds to $\overline{%
h}(M_{Pl})=2.1\div 3.5$) leads to the very large range of $\overline{v}%
(M_{Pl})$ (from 10$^{5}$ to 10$^{30}$ GeV, see Fig. 1) and $\overline{%
\lambda }(M_{Pl})$ (from 0.13 to 0.35, see Fig. 2) at the Planck scale for
the compactification scale $\mu _{c}=10^{16.75}$GeV in the case of one extra
dimension, $\delta =1$. Thus one can see, while the true fixed--point
corresponds to $\overline{\lambda }=\overline{h}=0$, relatively large values
of $\overline{\lambda }$ and $\overline{h}$ in the infrared region can be
obtained as well starting with the parameters at $M_{Pl}$ in a quite large
interval. Requiring that $\overline{\lambda }$ and $\overline{h}$ are within
the perturbative regime ($\frac{\overline{\lambda }^{2}}{4\pi }<1$ and $%
\frac{\overline{h}^{2}}{4\pi }<1$) and that the relations (\ref{23a})-(\ref
{23c}) hold for the whole interval between $M_{W}$ and $M_{Pl}$, our toy
model predicts the upper bounds on the physical masses of the scalar and
fermion: 
\begin{eqnarray}
m_{S} &\lesssim &73\mbox{ GeV}\ ,  \nonumber \\
m_{F} &\lesssim &100\mbox{ GeV}\ .  \label{25}
\end{eqnarray}
Finally, we have explicitly cheaked that the ratio $\frac{\overline{v}(%
\overline{v})}{\overline{v}(M_{Pl})}$ is actually stable under the variation
of the scalar VEV at Planck scale with $\overline{\lambda }(M_{Pl})$ and $%
\overline{h}(M_{Pl})$ fixed. Namely, for $\overline{\lambda }(M_{Pl})=0.2$
and $\overline{h}(M_{Pl})=3$ (and for $\mu _{c}=10^{16.75}$GeV, $\delta =1$
again) for the average value of the anomalous dimension between the scales $%
M_{Pl}$ and $v$ we obtain 
\begin{equation}
\left\langle A\right\rangle \equiv (\ln \frac{\overline{v}}{M_{pl}}%
)^{-1}\int_{\ln \frac{\overline{v}}{M_{pl}}}^{0}A_{\delta }(\Lambda )d(\ln 
\frac{\Lambda }{M_{Pl}})\approx \frac{2}{1+0.03\ln \frac{M_{Pl}}{\overline{v}%
(M_{Pl})}}.  \label{26}
\end{equation}
So, if $\overline{v}(M_{Pl})\approx M_{Pl}$, as it is naturally expected, $%
\left\langle A\right\rangle $ is close to 2, providing large stable
hierarchy $\frac{\overline{v}(\overline{v})}{\overline{v}(M_{Pl})}\approx
1.8\cdot 10^{-15}$. Thus, varying $\overline{v}(M_{Pl})$ by 10$\%$ around $%
10^{17}$GeV, we obtain $\left\langle A\right\rangle \approx 2.14\div 2.13$
and $\overline{v}(\overline{v})=157\div 190$ GeV.

It should be stressed that our solution to the gauge hierarchy problem does
not require the extra dimensions to be large. In fact, the hierarchy $\frac{%
\mu _{c}}{M_{Pl}}\sim 0.05\div 0.3$ is enough to get the desired values of $%
\overline{v}$ at low energies, even starting with naturally expected large
values of $\overline{v}$ at $M_{pl}$ ($\overline{v}(M_{pl})\sim M_{Pl}$).
This is demonstrated in Fig. 3, where we have plotted the evolution of $%
\overline{v}(\Lambda )$ for the case of $\delta =1,2,$ and 3 extra
dimensions starting from $\overline{v}(M_{pl})=10^{17}$GeV. Requiring that $%
\overline{v}(174$ GEV$)=174$ GeV and $\overline{\lambda }(174$ GEV$)=0.25$, $%
\overline{h}(174$ GEV$)=0.55$, we obtain $\mu _{c}=10^{16.75}$, 10$^{17.30}$%
, 10$^{17.51}$ GeV for $\delta =1,2,$ and 3, respectively. For the sake of
comparison we have also plotted the running of $\overline{v}(\Lambda )$ in
four dimensions ($\delta =0$). One can see that a familiar fine tuning of
parameters at the Planck scale is required in order to achieve the
phenomenologically acceptable values of $\overline{v}$ at low energies.

We conclude with the following comments. First, it is clear, that more
accurate calculations, related to an improved treatment of thresholds and
contributions beyond the LPA approximation and truncation (\ref{9}) of the
effective potential as well as higher loop corrections quantitatively alter
(perhaps quite significantly) our predictions for particle masses in (\ref
{25}), but qualitatively the behaviour of the parameters seems to remain
unchanged, thus providing us with a the natural solution of the gauge
hierarchy problem as discussed above. Second, while we have demonstrated our
mechanism for the solution of the gauge hierarchy problem on the simplified
model of Higgs--Yukawa interaction, we find no reason why it can not work in
the realistic models when a full set of the particles and forces of the SM
or its extensions will be included. Moreover, several examples of
nonsupersymmetric unification through extra dimensions have been recently
presented \cite{UN}, so we believe that a unified model without the gauge
hierarchy problem and consistent with present experimental data can be
constructed elsewhere.

We would like to thank Z. Berezhiani, J. Chkareuli, G. Dvali, M.
Gogberashvili, I. Gogoladze, T. Kobayashi and Z. Tavartkiladze for useful
discussions and comments.

\newpage {\large \textbf{Figure captions}}

\vspace{0.5 cm}

\textbf{Fig. 1}~ Fixed-point behaviour of the scalar field VEV in the case
of one extra dimension ($\delta $=1) with the compactification scale $\mu
_{c}=10^{16.75}$ GeV. \newline
\textbf{Fig. 2}~ Fixed-point behaviour of the scalar field self-interaction
coupling $\overline{\lambda }$ in the case of one extra dimension ($\delta $%
=1) with the compactification scale $\mu _{c}=10^{16.75}$ GeV. \newline
\textbf{Fig. 3}~ The typical running of the scalar field VEV in the case of $%
\delta $=1,2 and 3 extra dimensions with $\mu
_{c}=10^{16.75},10^{17.30},10^{17.51}$ GeV, respectively. The values of VEV $%
\overline{v}(M_{Pl})=10^{17}$ GeV and $\overline{v}(174GeV)=174$ GeV are
fixed. The running in four dimensions ($\delta $=0) with $\overline{v}%
(174GeV)=174$ GeV fixed is also plotted.

\newpage

\begin{figure}[tbp]
\begin{center}
\epsfig{file=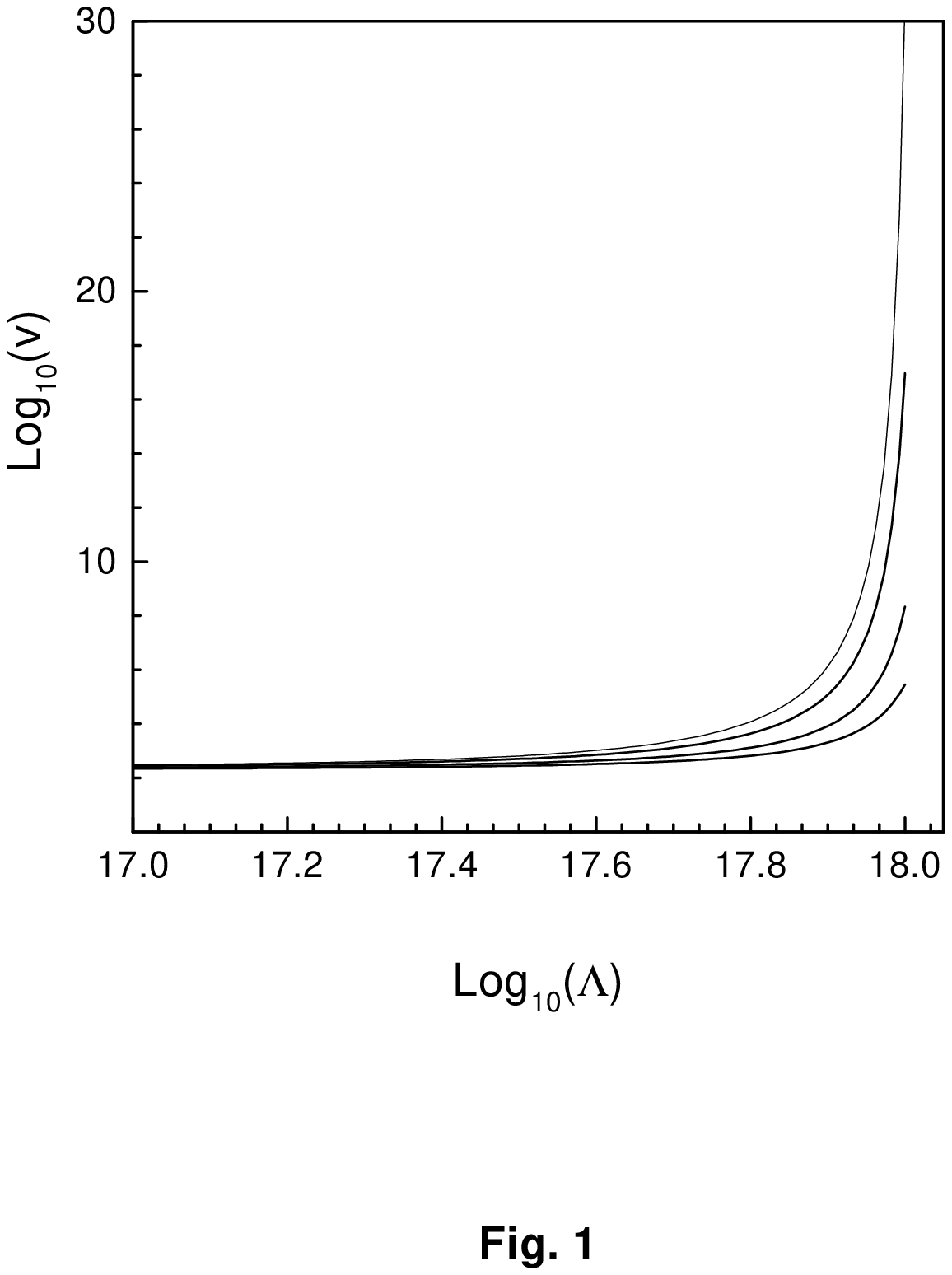} 
\end{center}
\caption{Fixed-point behaviour of the scalar field VEV in the case of one
extra dimension ($\delta$=1) with the compactification scale $%
\mu_c=10^{16.75}$ GeV.}
\end{figure}

\newpage

\begin{figure}[tbp]
\begin{center}
\epsfig{file=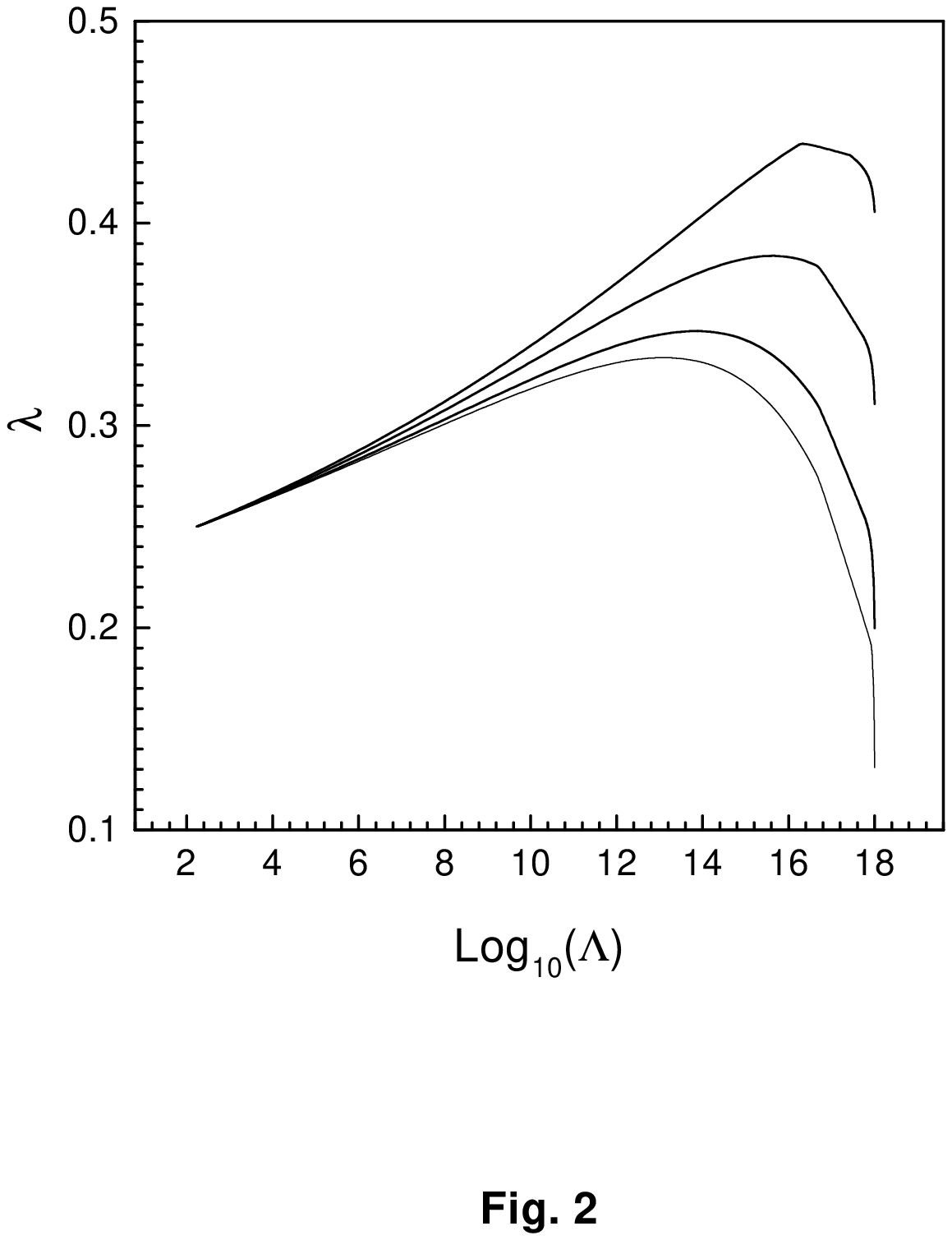} 
\end{center}
\caption{Fixed-point behaviour of the scalar field self-interaction coupling 
$\lambda$ in the case of one extra dimension ($\delta$=1) with the
compactification scale $\mu_c=10^{16.75}$ GeV.}
\end{figure}

\newpage

\begin{figure}[tbp]
\begin{center}
\epsfig{file=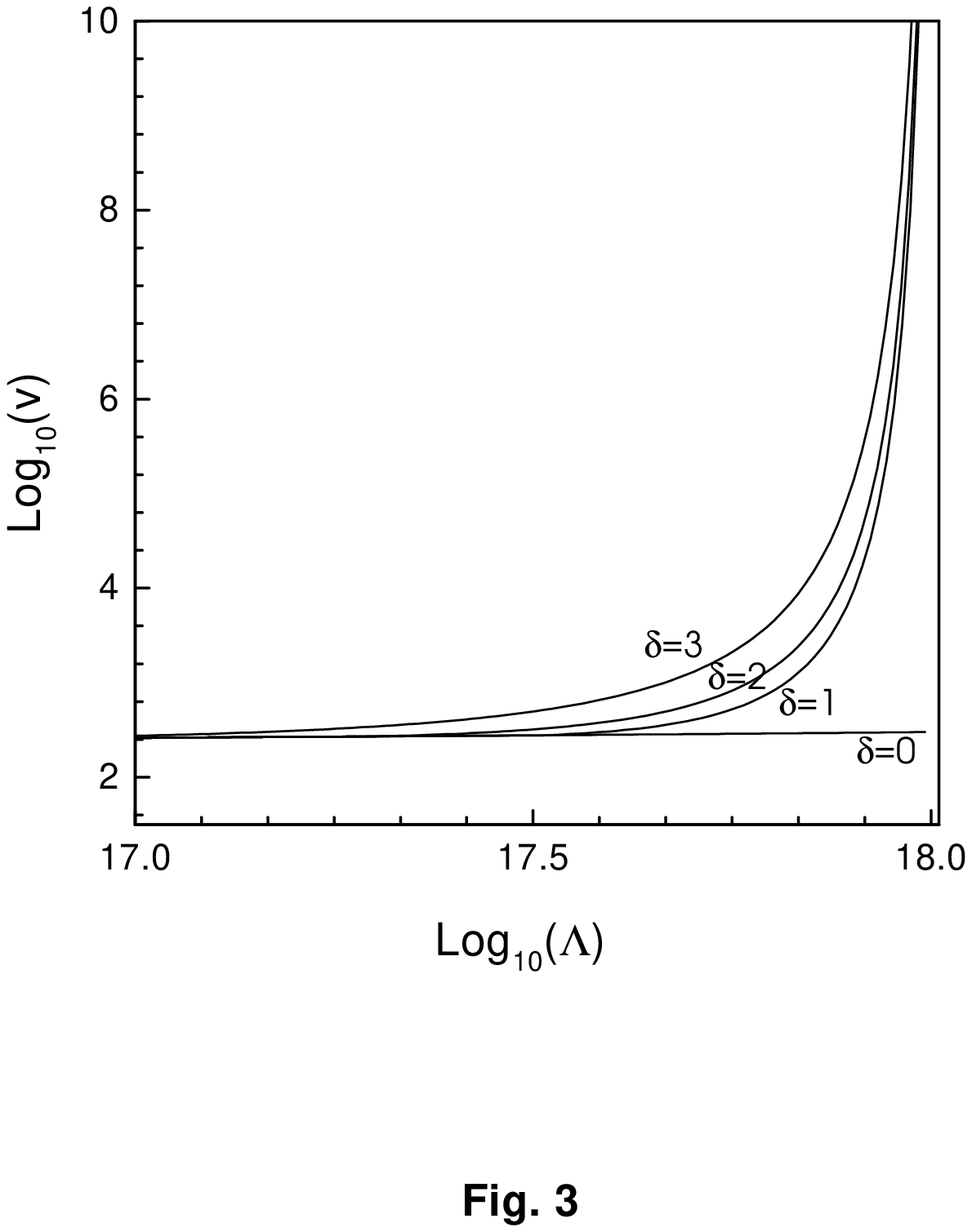} 
\end{center}
\caption{The typical running of the scalar field VEV in the case of $\delta$
=1,2 and 3 extra dimensions with $\mu_c=10^{16.75}, 10^{17.30}, 10^{17.51}$
GeV, respectively. The values of VEV $v(M_{Pl})=10^{17}$ GeV and $v(174
GeV)=174$ GeV are fixed. The running in four dimensions ($\delta$=0) with $%
v(174 GeV)=174$ GeV fixed is also plotted.}
\end{figure}

\end{document}